\documentclass[10pt,conference,a4paper]{ITKproc}
\usepackage[utf8]{inputenc}
\usepackage{amssymb, hyperref, caption, subcaption, svg, makecell, tabularx, pdflscape, xcolor, multirow, amsmath}

\title{Standardized Evaluation of Fetal Phonocardiography Processing Methods}
\author{\IEEEauthorblockN{Kristóf MÜLLER, Janka HATVANI, Márton Áron GODA, Miklós KOLLER}}

\begin{document}

\maketitle
\begin{abstract}
    \textit{Motivation.} Phonocardiography can give access to the fetal heart rate as well as direct heart sound data, and is entirely passive, using no radiation of any kind.
    \textit{Approach.} We discuss the currently available methods for fetal heart sound detection and heart rate estimation and compare them using a common benchmarking platform and a pre-selected testing dataset. Compared to previous reviews, we evaluated the discussed methods in a standardized manner for a fair comparison. Our tests included tolerance-based detection accuracy, error rates for label insertions, deletions, and substitutions, and statistical measures for heart rate mean square error. \textit{Results.} Based on our results, there is no definite best method that can achieve the highest scores in all of the tests, and simpler methods could perform comparably to more complex ones. The best model for first heart sound detection achieved 97.6\% F1-score, 97.4 positive predictive value, and 12.2$\pm$8.0 ms mean absolute error. In terms of second heart sound detection the best model had 91.4\% F1-score, 91.3 positive predictive value, and 17.3$\pm$12.2 ms mean absolute error. For fetal heart rate a 0.644 mean square error was achieved by the best method. \textit{Significance.} Our main conclusion is that further standardization is required in fetal heart rate and heart sound detection method evaluation. The tests and algorithm implementations are openly available at: \url{https://github.com/mulkr/standard-fpcg-evaluation}.
\end{abstract}


\section{Introduction}
Monitoring the health of the heart is an important medical topic, especially the early detection of congenital heart defects (CHD) for a better long-term outcome. The most common and reliable method for this diagnosis is echocardiography, an ultrasound technique \cite{Fetal-echocard-2019, Fetal-echocard-2004}. It is part of a more general genetic ultrasound screening at 20-22\textsuperscript{nd} weeks \cite{ISUOG-guidelines} and can reveal most mechanical and conductive defects in the heart. However, this technique requires a trained person to record and a medical expert to evaluate the results. Thus, these measurements necessitate hospital visits, which can cause excessive stress for both the mother and the fetus. In addition, exposing the fetus to high amounts of ultrasound should be limited.

Phonocardiography (PCG) offers an alternative measurement process, since the sounds produced by the heart carry potentially enough information for screening and diagnosis of fetal developmental problems and CHDs. Currently, it is not part of common medical practice outside of primary auscultation, but PCG signal processing techniques have gained attention recently. The main strengths of PCG lie in that the recording equipment is relatively cheap, as the hardware is less complex than that of a usual ultrasound machine, and PCG measurements do not require exposure to acoustic (or electromagnetic) radiation meaning that the signal acquisition is entirely passive.
In recent years the George B. Moody Challenge (formerly: PhysioNet Challenge), an international competition used PCG data for their problem statements \cite{Physionet2016, PhysioNet2022}. Both were aimed at developing a method for detecting abnormalities in the recordings, so that more remote and developing communities can get these benefits.

\begin{table}[t]
    \setlength{\tabcolsep}{4pt}
    \begin{center}
    \caption{Comparison of commonly used fetal cardiological measurement modalities}
    \begin{tabular}{l|l|l|l|l|l}
    \hline\hline
    \gape{Method} & Price category & Specialist & Harm & FHR & fHS \\ \hline
    \gape{US imaging} & High & Required & Potential* & \checkmark & $\times$ \\
    Doppler US & Middle-high & Recommended & Potential* & \checkmark & $\times$ \\
    \gape{fPCG} & Low-middle & Not required & No harm & \checkmark & \checkmark \\ \hline\hline
    \end{tabular}
    \end{center}
    *These are considered safe for short-term monitoring but there is no conclusive research to suggest their safety with long-term monitoring
    \label{tab:ctg-fpcg-comp}
\end{table}

\begin{figure}[t]
    \centering
    \includegraphics[width=0.9\columnwidth]{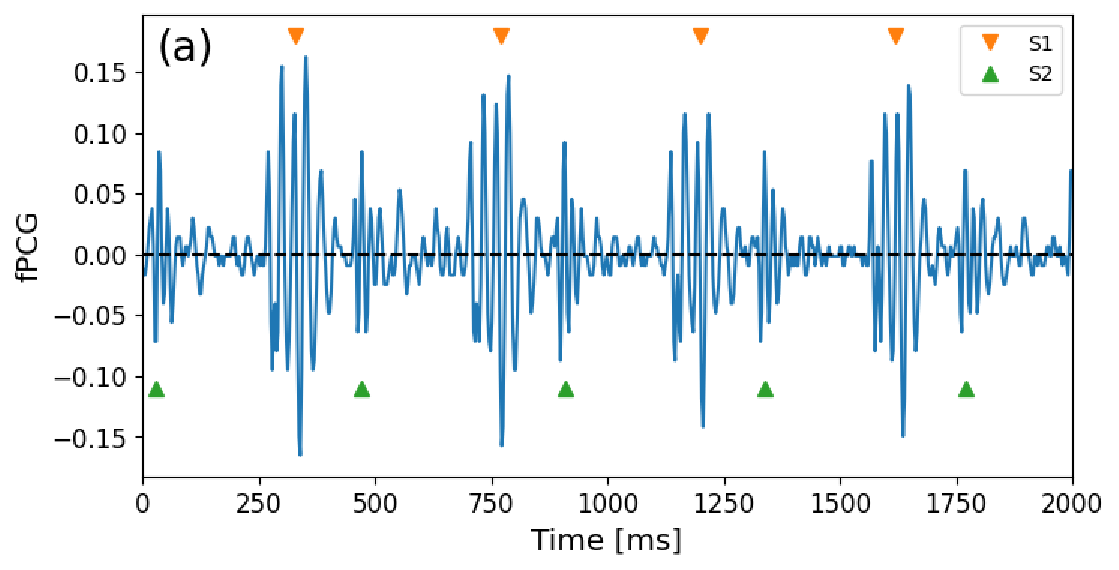}
    \includegraphics[width=0.8\columnwidth]{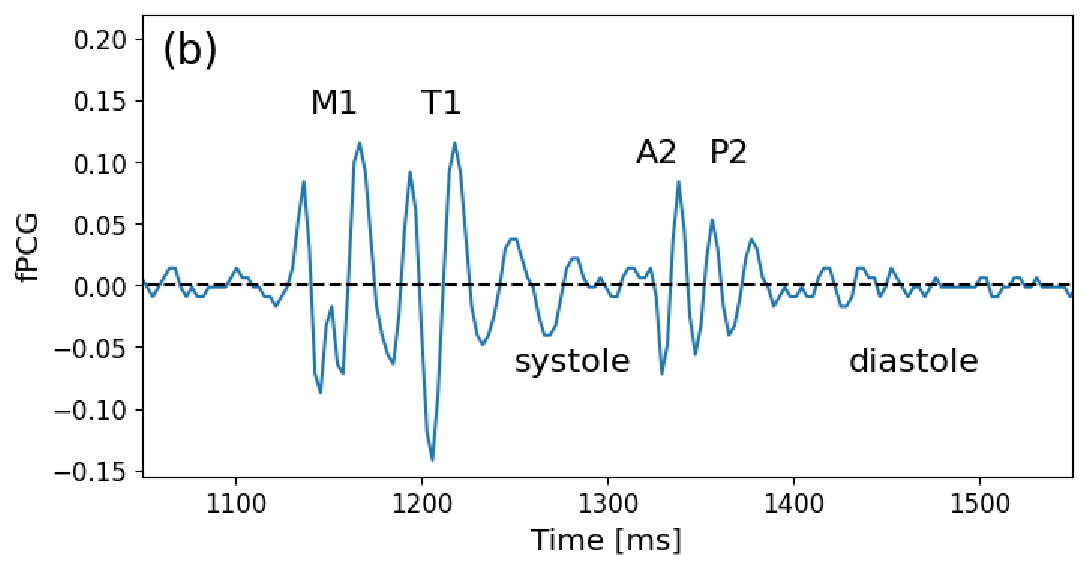}
    \caption{Samples from an fPCG signal. (a) Two second long sample with labeled heart sounds. (b) A single heart cycle with labeled heart valve components and heart cycle states.}
    \label{fig:fpcg}
\end{figure}

As PCG measurement is a passive process, it can be considered safer than ultrasound techniques for long-term fetal monitoring \cite{ultrasound-safety-2009, ultrasound-safety-2002, ultrasound-safety-2013}. During longer ultrasound measurements, especially with 3D or 4D recording techniques the temperature of the measured tissue can increase, which is dangerous for less developed fetuses. Additionally, cavitation can also occur which can cause tissue damage, although the mechanical index (the probability of cavitation) is strictly controlled in medical devices. Since fetal PCG (fPCG) can be safely and easily performed, it increases the amount of usable data which can lead to more robust detection of CHDs, however, a significant drawback of fPCG is that currently the earliest measurements are only possible after the 24\textsuperscript{th} week. Nevertheless, fPCG opens the possibility of a longer monitoring period over several weeks, which can also lead to additional information for monitoring fetal development, such as fetal lung maturity or the risk of preterm birth \cite{fetal-lung-maturity,preterm-birth-risk}.

Previous reviews have discussed multiple methodologies for fPCG processing, such as fetal heart rate estimation, signal denoising, and heart sound detection \cite{method-review,techniques-review} and acquisition devices \cite{AcqDevices-review,fpcg-ctg-comparative}. However, the mentioned works are not always comparable since the datasets and evaluation metrics are not shared between the methods. Our work aims to further discuss these approaches mainly by evaluating each algorithm on a shared dataset and with the same accuracy measures.

\section{Biological background}

The motion of the heart intrinsically produces certain physical phenomena, such as the mechanical vibrations which are usually caused by the opening and closure of the heart valves, as well as potential turbulent flow of the blood. The heart has two pairs of valves on each side to regulate blood flow during the heart cycle. The mitral and tricuspidal valves control the blood flowing into the ventricles, while the pulmonic and aortic valves regulate the blood leaving the ventricles into the arteries. In healthy adults the closure of the valve pairs happen synchronously, causing the first and the second heart sounds, called S1 and S2. A third heart sound (S3) can also be heard in pediatric cases after S2 and at the beginning of diastole, the presence of S3 in adults is considered to be pathological. The fourth heart sound (S4) appears near the onset of S1 and is always considered pathological. The main S1 and S2 sounds can also be \textit{split}, where the small delay between the valve impulses increases. This can happen while breathing, however, the amount of delay and its change have to be monitored, since it can be a sign of certain heart defects. In the S1-S2 and S2-S1 intervals, the systole and diastole, respectively, no noise should be heard during auscultation. However, due to some valve defect or other abnormality, the bloodflow can become turbulent or even retrograde, which causes distinctive noises in the PCG, called murmurs or clicks.

Current standard fetal heart monitoring does not measure the raw PCG for later analysis, only the fetal heart rate (FHR) is considered. FHR is usually measured with Doppler ultrasound from the 6-8\textsuperscript{th} weeks and after the 37-38\textsuperscript{th} week it is measured along with womb contractions. Devices which combine these modalities are known as cardiotocographs (CTGs). CTG measurements are also carried out during intrapartum where there are any underlying risks, to measure the health of the to be born baby. Analysis of the FHR curve is subject to several international standards \cite{figo}, however, it was found that statistically CTG analysis does not increase general well being of the infant. Two Cochrane reviews in 2015 and 2017 were done to evaluate the consensus on the benefits of antepartum and intrapartum CTG monitoring \cite{CTG-continuous, CTG-antenatal} and both found a lack of good quality research, mainly due to several biases introduced. The 2017 study was focused on finding a connection between continuous CTG and improved well-being of the baby. Based on the reviewed works the findings in most cases had very low evidence, except where computerised CTG evaluation was used, which showed promising results. The 2015 review about antepartum monitoring observed similarly low quality research. The conclusion for both articles associated CTG with a reduction in neonatal seizures, but infant mortality, and standard measures of neonatal health showed low or no significant connection. Additionally, they observed that the use of CTG increased the number of cesarean sections and instrumental vaginal births. From these results the authors concluded that more and higher quality research has to be done in this field.

In contrast to CTG, fPCG provides a more detailed signal, which can not only be used to determine FHR but with other processing methods it can be used to detect phenomena not observable with CTG. These include heart sound splitting and murmurs \cite{Muller-split,Balogh-murmur,fetal-murmur-chd-screening,fetal-murmur-computer-method}, which can serve as a good basis for congenital heart defect screening.

\begin{figure}[t]
    \centering
    \includegraphics[width=0.9\columnwidth]{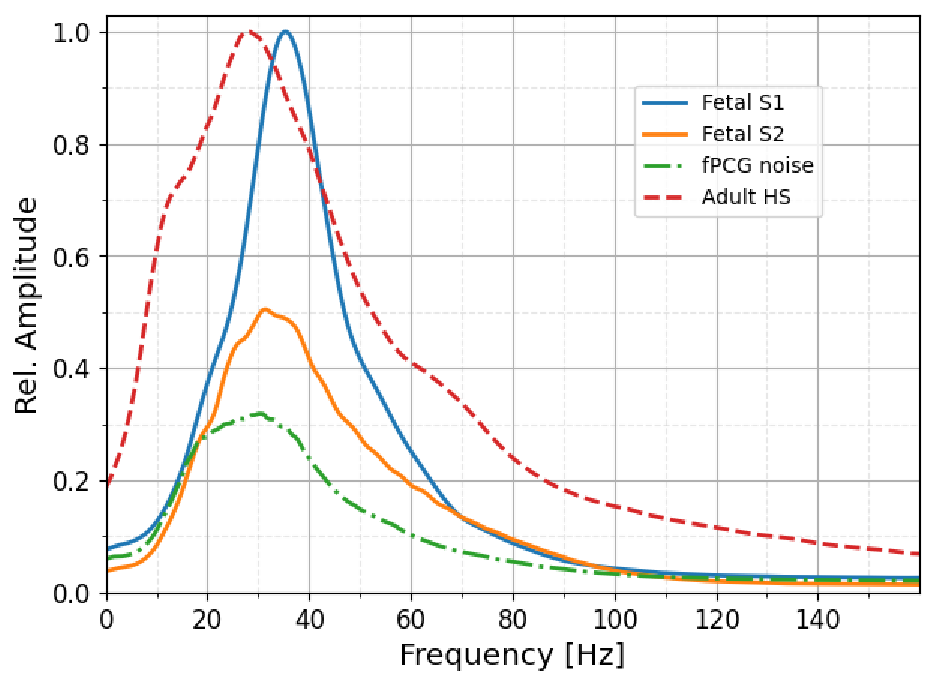}
    \caption{Comparison of average spectra of adult and fetal heart sounds (HS). Adult heart sounds extracted from: \cite{authhsdb,UHAHSDB-2011,UHAHSDB-2013}. Annotated fetal heart sounds from the 50 one-minute-long records used by \cite{muller2024pypcg} which were selected from \cite{Kovacs2011}}
    \label{fig:hs-freq}
\end{figure}

\section{Fetal PCG processing challenges}
\subsection{Maternal factors}
Different types of maternal influence are the main components which need to be addressed for accurate fPCG measurements. The most important detail is that the fetal heart sounds (fHS) are recorded through the abdomen and the womb. The amniotic fluid, the muscle walls of the womb and other internal tissue between the recording device and the fetus serve as a low-pass filter, thus high-frequency data from the fetal heart cannot be easily recorded \cite{womb-lowpass}. The other significant factor is that the maternal heart is a more developed organ and can produce louder heart sounds than the fetal heart \cite{Kovacs2011-possibilities}. If the recording sensor is placed incorrectly, then these sounds can appear in the final signal. These have to be filtered out when processing the fPCG because they can cause erroneous detections. Filtering is usually done based on spectral information of the sounds or their observed periodicity, since the maternal heart sounds are lower in frequency and amplitude and the fetal heart rate is significantly higher than the resting maternal heart rate. For a comparison between the spectra of an adult heart signal and fetal heart sounds see Figure \ref{fig:hs-freq}. The other maternal factors include the respiratory sounds and noise produced by the digestive system. These do not cause a serious problem in processing because they can be reduced relatively well. Breathing in the signal occupies a lower frequency range therefore with a high-pass filter this can be filtered out. Digestive sounds are temporally limited thus a long enough recording would contain segments where these are not occurring and analysis can be done only using that information.

\begin{table*}[t]
\centering
\caption{Summary of open source fetal phonocardiography (fPCG) databases}
\begin{tabular}{l|l|l|l|l|l|l}
\hline\hline
\gape{Publication} & Data origin     & Gestational age        & Signal properties        & Average length & Annotations & Metadata \\ \hline
Cesarelli \textit{et al.} (2012) \cite{fpcg-simul} & Simulated       & \gape{N/A}                    & \gape{\makecell{1000 Hz,\\16 bit int}}     & 8 minutes           & - & - \\
Cesarelli \textit{et al.} (2012) \cite{fpcg-simul} & 26 pregnancies  & \gape{31-40 weeks}            & \gape{\makecell{333 Hz,\\8 bit unsigned int}}       & 20 minutes          & - & \makecell{Gestational\\week}\\
Samieinasab \& Sameni (2015) \cite{shiraz} & 112 pregnancies & \gape{30-40 weeks} & \gape{\makecell{8000 Hz/16 kHz,\\16 bit int}}    & 85 seconds          & FHR & multiple\\
Bhaskaran \& Arora (2022) \cite{indian-institute-data} & 99 pregnancies  & \gape{30-40 weeks}            & \gape{\makecell{2000 Hz,\\64 bit float}}   & 9 minutes           & Average FHR & multiple\\ \hline\hline
\end{tabular}
\label{tab:fpcg-summary}
\end{table*}

\subsection{Fetal factors}
Fetal factors are more manageable and can be used to extract additional information about the fetus. The first one being the position of the fetus, which can cause the heart sounds to have a lower amplitude in certain auscultation locations \cite{FHR-zahorian}. The other main fetal factor is caused by different movements of the fetus. These movements can be the general activity, such as kicking or hiccups, but respiratory movements are also possible. This was previously demonstrated to be detectable with fPCG \cite{FetalBreathing}, and can be possibly linked to the development of the fetal respiratory system, thus by measuring this the risks of preterm births could be better evaluated.

\section{Openly Available Fetal Data}
The greatest limitation in fPCG research currently is the lack of available heart sound labeled data. Most open datasets focus on fetal heart rate calculation and do not provide individual heart sound labels or include information about pathologies. In this section we will detail the most commonly used open datasets, of which a summary can also be seen in Table \ref{tab:fpcg-summary}.

Based on a pilot study by measuring physiological singleton pregnancies Cesarelli \textit{et al.} developed a dataset comprised of simulated signals (\textit{simfpcgdb}) \cite{fpcg-simul}. The simulation had three main steps: FHR simulation,  fHS generation, and adding noise. The simulated FHR was parameterized to be as realistic as possible, with accelerations and decelerations as well as more extreme cases of bradycardia and tachycardia. The heart sounds were modelled as Gaussian-modulated sinusoidal pulses, which was previously shown to be an accurate model \cite{artificial-womb}. They also observed that the power spectral density (PSD) of the heart sounds was similar to a Gaussian curve. By fitting a Gaussian to the heart sound PSDs and setting the other parameters based on the pilot study their model of FHS was more biologically accurate. The authors also simulated several levels of additive noise in the data. Based on the pilot study and other sources five types of noise were considered: maternal heart sounds (modelled similarly to FHS), other internal noises (maternal organs, fetal movements) with an average low frequency, environment with high frequencies, general white noise originating from the amplifier, and limited duration impulses which saturate the signal with noise (moving the sensor, coughing). These noise types were fine-tuned based on observations and added to the signal. As a continuation of this project, the recordings created in the pilot study were also made public via PhysioNet, named ``Fetal PCG Database" (\textit{fpcgdb}). This contains 26 recordings made with a Fetaphon device by Pentavox. The data was collected from singleton pregnancies between 31 and 40 weeks of gestation. The signals were recorded at 333 Hz sampling rate and an 8 bit analog to digital converter.

A commonly used open fPCG recording dataset containing real signals is the Shiraz University Fetal Heart Sounds Database developed by Samieinasab and Sameni (\textit{sufhsdb}) \cite{shiraz}. Their database was built by recording 110 women between their 30th and 40th week of gestation with the JABES electronic stethoscope using its ``wideband" mode. Simultaneous FHR data was recorded with CTG so that comparison with fPCG-based FHR calculation can be performed. The signals were digitized with an 8 kHz sampling rate and 16 bits of precision. The authors used this data for their further work to describe a denoising method based on single channel blind source separation. Their method made use of empirical mode decomposition (EMD) \cite{EMD-alg} and nonnegative matrix factorization (NMF) \cite{NMF-alg}. After applying the EMD, each resulting intrinsic mode function (IMF) was transformed with short-time Fourier transform (STFT) and the NMF was applied to these results. After an inverse Fourier transform, a clustering step was used to separate the fPCG, respiratory noise, and Gaussian noise.

Another open database with real signals is the one created and used by Bhaskaran \textit{et al.}, also known as the ``Indian Institute of Science dataset" (\textit{iiscfhsdb}) \cite{indian-institute-data}. Similarly this was also compiled for fPCG based FHR calculation. The recordings were collected from 99 pregnant women between 30 and 40 weeks of gestation with no known maternal complications. The fPCG signals were recorded with an electronic stethoscope, 2 kHz sampling frequency, and gain of 500. Out of these signals 10 were recorded in a pilot study, 15 used a notch filter at 50 Hz to reduce the power supply interference, and the remaining 74 did not use this filter because the power supply was a battery. The FHR reference was determined manually when possible; only 60 recordings were deemed acceptable for manual annotation. The recordings were preprocessed with a 4th-order Butterworth band-pass filter. Three possible bands were chosen, 10-40 Hz, 20-50 Hz, or 30-60 Hz. These proved to be acceptable for all 60 signals. Determining the FHR was done in 4 second time-windows by marking each S1 sound. This annotation utilized two observers and the calculated FHR was rejected if the inter-observer difference was higher than 10 beats per minute (bpm). The authors also describe an automatic FHR calculation method based on the Hilbert envelope and autocorrelation. An extra \textit{peak validation} step was also described which ensured the accuracy of the automatic method, if this validation failed then the given segment did not receive an FHR estimate and a gap was inserted in the FHR curve.

\begin{landscape}
\begin{table}[p]
\begin{center}
\caption{Summary of the discussed fetal heart sound (fHS) detection methods}
\begin{tabular}{l|l|l|l|l|l|l}
\hline\hline
\thead{fHS detection method} & \thead{Dataset} & \thead{Preprocessing} & \thead{Detection} & \thead{Validation} & \thead{Accuracy measures} & \thead{Implementation} \\ \hline
Chen \textit{et al.} (2006) \cite{FHR-rms} & Internal & \gape{\makecell{Analog LP filter (110 Hz),\\Digital HP filter (35 Hz),\\Simplified Spectral Subtraction}} & \gape{\makecell{RMS, threshold local maximum,\\peak merge, amplitude regularity}} & CTG & Visual assessment & \checkmark\textsuperscript{1} \\
Schmidt \textit{et al.} (2010) \cite{Schmidt_2010} & Internal & \gape{\makecell{Band pass filter (25-400 Hz),\\Spike removal,\\Envelogram calculation}} & \gape{\makecell{Hidden semi-Markov model}} & Manual labels & Sensitivity, PPV & \checkmark \\
Balogh and Kovács (2011) \cite{Balogh-murmur} & Internal & - & \gape{\makecell{Heuristic processing with\\multiple difference levels \\with smaller window sizes}} & - & - & \checkmark\textsuperscript{1} \\
Cesarelli \textit{et al.} (2012) \cite{fpcg-simul} & \textit{simfpcgdb} & \gape{\makecell{Band pass filter (34-54 Hz),\\Teager energy operator}} & \gape{\makecell{Local maxima finding}} & Simulation ground truth & \gape{\makecell{Accuracy,\\Percentage of missed beats}} & \checkmark\textsuperscript{1} \\
Springer \textit{et al.} (2016) \cite{HSMM} & Internal & \gape{\makecell{Band pass filter (25-400 Hz),\\Spike removal,\\Envelogram calculation}} & \gape{\makecell{Hidden semi-Markov model\\with logistic regression}} & ECG & \gape{\makecell{Sensitivity, Accuracy,\\PPV, F1-score}} & \checkmark \\
Koutsiana \textit{et al.} (2017) \cite{Koutsiana2017} & \gape{\makecell{Internal simulated,\\\textit{simfpcgdb}}} & \gape{\makecell{Wavelet decomposition\\detail component selection (\textit{db4})}} & \gape{\makecell{Fractal dimension,\\Peak peeling algorithm}} & Database ground truth & \gape{\makecell{Efficiency index,\\Correct detection percentages}} & - \\
Renna \textit{et al.} (2019) \cite{Renna-UNET} & PhysioNet 2016 & \gape{\makecell{Band pass filter (25-400 Hz),\\Spike removal,\\Envelogram calculation}} & \gape{\makecell{U-net like CNN\\Temporal modeling}} & Database ground truth & Sensitivity, PPV, Accuracy & \checkmark\textsuperscript{2} \\
Tomassini \textit{et al.} (2020) \cite{AdvFPCG} & \gape{\makecell{\textit{sufhsdb},\\ \textit{simfpcgdb}}} & \gape{\makecell{Band pass filter (20-120 Hz),\\Wavelet denoising (coif4)}} & \gape{\makecell{Continuous wavelet transform,\\Peak finding with timing rules}} & \gape{\makecell{Manual labels,\\database FHR}} & \gape{\makecell{Number of detections,\\fHS timing intervals,\\FHR difference}} & - \\
Vican \textit{et al.} (2021) \cite{Vican-EMD} & Internal & \gape{\makecell{Downsampling (2 kHz),\\Bandpass filter (50-150 Hz),\\Empirical mode decomposition}} & \gape{\makecell{Support vector classifier\\Random forest\\Multilayer perceptron}} & CTG & S1 classification accuracy & \checkmark \\
Almadani \textit{et al.} (2023) \cite{unet-transformer} & Internal & - & \gape{\makecell{Transformer neural network,\\U-Net neural network}} & fetal ECG & \gape{\makecell{FHR difference,\\visual inspection}} & - \\
Müller \textit{et al.} (2024) \cite{muller2024pypcg} & Internal & \gape{\makecell{Band pass filter (15-55 Hz),\\Spike removal,\\Envelogram calculation}} & \gape{\makecell{Hidden semi-Markov model\\with logistic regression\\and optimized timing parameters}} & Manual labels & \gape{\makecell{PPV, F1-score,\\MAE, Error rate,\\Score-vs-Tolerance}} & \checkmark \\
Kong \textit{et al.} (2024) \cite{hybridized-classifier} & \makecell{PhysioNet Challenge 2016,\\Springer data,\\custom fetal PCG simulation} & Fourier Synchrosqueezed Transform & \gape{\makecell{Transformer neural network,\\Random forest with XGBoost,\\Hybridized decision rule}} & Database ground truth & \gape{\makecell{Sensitivity, specificity,\\F1-score, accuracy}} & Partial \\ \hline\hline
\end{tabular}
\label{tab:fhs-summary}
\end{center}
\textsuperscript{1}: Reimplemented by us, based on original paper for this evaluation, \textsuperscript{2}: Reimplemented by \cite{Unet-impl}
\end{table}
\end{landscape}

\section{Heart sound detection}
In this section predominantly fetal heart sound (fHS) detection methods are described, summaries of these methods can be seen in Figure \ref{fig:fhr-flow} and Table \ref{tab:fhs-summary}.

\subsection{Noise filtering}
Almost all PCG processing methods, not exclusive to heart sound detection but also heart rate estimation, start with a noise filtering step to improve the signal quality and also to limit the possible frequency band to that which is physiologically probable. This is usually achieved with analog or digital filtering, but in some cases  decomposition based denoising was used, although these are not mutually exclusive procedures. In our set of examined methods, filtering methods show an interesting diversity mostly in the frequency band chosen, but also in filtering types (band-pass or low-pass filters, wavelet or empirical mode decomposition). This is illustrated in the third column of Table \ref{tab:fhs-summary}.

In terms of frequencies chosen for filtering we can observe that most methods are concerned with a band between 15-400 Hz. This is due to the fact that the heart sounds and especially the fetal heart sounds are very limited in bandwidth, according to \cite{NAGEL, artificial-womb, womb-lowpass}. Mathematically these digital filters can be described with their transfer function ($H(z)$) as
\begin{equation}
    H(z)=\frac{B(z)}{A(z)} = \frac{\sum_{k=0}^Nb_kz^{-k}}{\sum_{k=0}^Ma_kz^{-k}},
\end{equation}
where $z$ is a general complex number, $a_k$ and $b_k$ are the coefficients of the numerator and denominator polynomials (degree $N$ and $M$) respectively. Also by convention $a_0$ is set to 1 without the loss of generality. If $M$ is set to 0, meaning the denominator is 1 a \textit{finite impulse response} (FIR) filter is described, otherwise, it is an \textit{infinite impulse response} (IIR) filter. Additional behavior of the filter can also be designed by setting the $B$ and $A$ polynomials to specific forms, for example by using the Butterworth design, the filter will have maximally flat pass and stop frequency bands.
\subsection{Envelope and energy}
Heart sounds in a good quality PCG signal can be easily detected with the signal envelope or an energy calculation. In most cases an envelope of the signal is derived using the Hilbert transform. This transformation ($\mathcal{H}$), for a signal $x$ it can be defined as a multiplier operator:
\begin{equation}
    \mathcal{F}\left[\mathcal{H}(x)\right](\omega)=-i\,\mathrm{sgn}(\omega)\cdot\mathcal{F}[x](\omega)\;,
\end{equation}
where $\mathcal{F}$ is the Fourier transform, $\mathrm{sgn}$ it the signum function, and $i$ is the imaginary unit. The Hilbert transform this way is effectively a phase shift of the signal by $\frac\pi2$. Using this, the analytical signal is defined as
\begin{equation}
    x_a(t) = x(t)+i\, \mathcal{H}(x(t)),
\end{equation}
thus the analytical signal is a complex valued function. Rewriting $x_a$ into polar form we get
\begin{equation}
    x_a(t) = x_m(t)\mathrm{e}^{i\pi x_\phi(t)},
\end{equation}
where $x_\phi$ is the instantaneous phase, and $x_m$ is the instantaneous amplitude also called as the Hilbert envelope.
An extension of the Hilbert envelope called the homomorphic envelope is also extensively used because it results in a smoother envelope signal due to the homomorphic (nonlinear) filtering, defined as
\begin{equation}
    x_h(t) = \mathrm{exp} (\,\mathrm{LPF} (\,\log ( x_m(t)))),
\end{equation}
where $\mathrm{LPF}$ is a low-pass filter, $\mathrm{exp}$ and $\log$ is exponentiation and logarithm with a given base (conventionally the natural base), respectively. The motivation behind this formula is that conventional filters are only capable of filtering additive noise. By taking the logarithm the possible multiplicative noise becomes additive and can be removed with a filter, then the signal is transformed back using exponentiation.
\begin{itemize}
    \item Envelogram calculation in \cite{Schmidt_2010, HSMM, Renna-UNET, muller2024pypcg} makes use of both the Hilbert and the homomorphic envelope as well as other envelopes. The internal low-pass filter in the homomorphic envelope uses a first-order Butterworth design with 8 Hz as the cutoff frequency. The envelogram is treated as different features for a given time location which are used for estimating a probabilistic model to best fit these observed features. This model will be further detailed in another section.
\end{itemize}
The local energy of the signal can also be used as an envelope-like signal, this can be calculated in both time-domain and frequency-domain as per Parseval's theorem. Stating that
\begin{equation}
    \label{eq:parseval}
    \int^\infty_{-\infty}|x(t)|^2dt=\frac1{2\pi}\int^\infty_{-\infty}|X(\omega)|^2d\omega\;,
\end{equation}
where $X(\omega)$ is the Fourier transform of $x(t)$.
Calculating the energy of a PCG signal is usually not calculated for the entire frequency range just a sub-band. Using a spectrogram to calculate the energy a time varying energy is obtained, since the spectrogram is a time-frequency representation.
Arguably the simplest and fastest way to calculate an energy estimate in time-domain is by calculating its root-mean square (RMS). The RMS of a signal $x$ is calculated as
\begin{equation}
    \mathrm{RMS}(x) = \sqrt{\frac1L\sum_{t=0}^L x(t)^2}\;,
\end{equation}
This formula results in a single value for a signal, however, by calculating the RMS in shorter time windows across the signal, a time varying RMS is achieved.
Using the Teager (or Teager-Kaiser) energy operator the signal energy can be estimated in an efficient way. This operator is defined for discrete signals ($x(n)$) as
\begin{equation}
    \Psi[x(n)] = x^2(n)-x(n-1) \times x(n+1) \;.
\end{equation}
\begin{itemize}
    \item Chen \textit{et al.} \cite{FHR-rms} used the RMS envelope in their real-time processing method to detect the S1 heart sounds. By merging local maxima labels based on a temporal and amplitude criteria approximate detections were achieved. These were then further filtered for physiologically plausible heart rate values based on their beat-to-beat time.
    \item For envelogram calculation the energy is calculated using the mean power spectral density between 40-60 Hz for the input signal \cite{Schmidt_2010, HSMM, Renna-UNET, muller2024pypcg}.
    \item In their article describing the fPCG simulation process Cesarelli \textit{et al.} also propose an S1 heart sound detection method as a proof of concept \cite{fpcg-simul}. In it the Teager-Kaiser energy is calculated and a local maxima finding algorithm is employed to detect heart sound candidates. The local maxima search is done in a time window based on the last identified heart sound and the average time difference between the last eight detections to ensure regularity. The energies of the previous beats are also used to determine a threshold value as half of the average energies. In a given window potentially multiple detection candidates can be found, the best fitting candidate is selected from these based on which is nearest to the location predicted by the average difference. Before this process a short training phase is used to set the initial timing and amplitude parameters.
    \item Earlier our research team demonstrated a simple S1 heart sound detection using the Teager-Kaiser energy for a preliminary split detection method \cite{Muller-split}. Their method uses a local maximum search using a predefined minimal peak distance, which was chosen to be longer than the average systolic interval to reduce the S2 detections but short enough so that high heart rates are also detected. From these candidate peak locations the statistical outliers were removed based on their peak-to-peak time and energy amplitude.
\end{itemize}
\subsection{Decomposition}
With decomposition methods the signal is separated into other signals based on certain rules. Most widely used are the wavelet decomposition (or wavelet transform) and the empirical mode decomposition (EMD) with their extensions. As mentioned previously these can be used for denoising by selectively applying some sort of filter to the components and reconstructing the signal or used as features for later classification or segmentation steps.

There exists two main types of wavelet transform: continuous (CWT) and discrete (DWT) wavelet transforms. The CWT is defined as
\begin{equation}
    W(s,u) = \frac1{\sqrt s} \int_{-\infty}^{+\infty}x(t) \ \psi^* \left( \frac{t-u}s \right)dt \;,
\end{equation}
where $s$ and $u$ are called the scale and delay parameters, $\psi$ is a so called \textit{mother wavelet} which satisfies certain properties, and the $^*$ operator means the complex conjugate. DWT, the discrete counterpart of the CWT restricts the possible range of the $a$ and $b$ parameters to discrete values, and is usually realized as a cascade process with low and high-pass filter pairs and downsampling stages. The result after a single stage is a \textit{detail} and an \textit{approximation} component which come from the high-pass and the low-pass branches, respectively. By further processing the approximation component, another detail and approximation components could be extracted, this is the cascading part of the process. The number of processing stages is called the \textit{decomposition level}, and each filter pair is designed carefully to satisfy similar properties to the previous mother wavelet.

Empirical mode decomposition (EMD) aims to extract information which can be interpreted as coming from a physical source. This technique can be intuitively thought of as separating the different superimposed oscillations originating from an object, which are called \textit{intrinsic mode functions} (IMFs). The EMD process can be described with the following steps:
\begin{enumerate}
    \item extract the upper and lower envelopes
    \item average the envelopes and subtract from signal
    \item save difference as an IMF
    \item repeat 1-3 with the last extracted IMF until given number of IMFs are extracted or the last IMF energy is below a threshold
\end{enumerate}
Additionally to denoising, CTW/DWT and EMD can be used to extract more features from a PCG signal, to improve the accuracy of detection or classification methods.
\begin{itemize}
    \item With selecting a given detail level, the absolute value of DWT (usually rbio3.9 family) is used in envelogram calculation \cite{Schmidt_2010, HSMM, Renna-UNET, muller2024pypcg}.
    \item Based on the work done by Vican \textit{et al.} \cite{Vican-EMD}, EMD is shown to increase the classification accuracy of heart sounds. Their process included extracting several features from fPCG recordings in a windowing manner to build a feature dataset. Features included several statistical measures and spectral features which were calculated for the preprocessed signal as well as the IMFs. For each fPCG signal, a CTG recording was also recorded as a ground truth for the S1 locations. These locations were used to separate the feature windows into positive and negative classes based on if they contained an S1 label. Multiple machine learning models were trained on this data, including support vector machines, random forests, and multi-layer perceptrons. Their results suggested that including IMFs improved the accuracy of the classification models.
    \item Tomassini \textit{et al.} proposed an extension to a previous heart sound detection method, called AdvFPCG-Delineator \cite{AdvFPCG, pcg-delin}. The original PCG-Delineator algorithm uses wavelet decomposition to denoise the input signals, but it uses an amplitude threshold along with certain timing parameters to detect fHS and differentiate between them. The extended version of this method first calculates a CWT scalogram using coif4 wavelet family, then the detection is performed using the scalogram. The PCG-Delineator first detects all possible S1 peaks, these are then filtered based on their peak-to-peak time, which needs to be at least 300 ms. The description of the original method includes a backsearching step to correct for S1 sounds which were not detected. S2 detection is performed similarly, with the additional constraints that these peaks need to be between two S1 peaks, and has to have at least 100 ms delay for the previous and at most 200 ms delay for the next S1 peak.
    \item Koutsiana \textit{et al.} used the db4 wavelet family DWT to decompose the input PCG \cite{Koutsiana2017}. Their method selected a given detail component for further processing (detailed in the next section). The selection was based on complex rules which involved first calculating the sum of the average energy content of the first $\lambda$ components, and the ratio of these compared to all components. Using these ``explained energy" ratios ($\eta_\lambda$), their first and second derivatives, and a threshold parameter $p$, a set of detail components were selected. These were then used as the inputs of the next step in their process.
\end{itemize}
\subsection{Complexity}
Measuring the local complexity can be used to classify biomedical signals and to detect events in them. Complexity can be described by the previously mentioned Shannon entropy measure, but other non-linear analysis methods such as fractal dimensions can also be used. The most widely used fractal dimension (FD) calculation method for signals is the Katz fractal dimension \cite{KFD}, which is defined as
\begin{equation}
    D=\frac{\log_{10}(L/a)}{\log_{10}(d/a)} \;,
\end{equation}
where $L$ and $a$ are the sum and average Euclidean distances of the samples in the signal, respectively, and $d$ is the maximum distance between the first sample and a given sample.
\begin{itemize}
    \item The previously mentioned method by Koutsiana \textit{et al.} processes the selected wavelet detail components by calculating the Katz fractal dimension (KFD) in a windowing manner to gather a local complexity measure. For peak detection they use an algorithm called fractal dimension peak peeling, previously proposed by Hadjileontiadis \cite{FD-peakpeeling}. The peak peeling algorithm first creates soft-thresholded signals based on the mean and standard deviation of the FD signals, where the threshold value is the sum of the mean and the standard deviation, additionally the values are reduced to one instead of zero if they are below the threshold value. Using this thresholded signal an intermediary signal is constructed using its difference with the original FD, then adding the mean value. The thresholding and intermediary signal calculation is repeated until the average energy difference of the previous and the current intermediary signal is less than a preset small value. The final soft-thresholded signal is then used to determine the heart sound locations using a hard threshold, which gives a binary mask to separate the heart sounds. S1 and S2 sounds were separated based on the delay between the detections and the assumption that in physiological cases the systolic interval is shorter than the diastolic interval.
\end{itemize}
\subsection{Machine learning}
Currently the most widely used method to segment the heart cycle in phonocardiographic signals is based on hidden Markov models (HMM). Where each hidden state corresponds to a heart cycle state, such as S1, systole, S2, and diastole. A HMM such as this can be described by its state transition matrix, which contains the probabilities of transitioning into another state based on the current state. A model like this follows the Markov property, which states that the probability of transitioning to a new state is only based on the current state. Based on the estimation by Schmidt \textit{et al.} \cite{Schmidt_2010}, a matrix for this process is
\begin{equation}
\mathbf{A}_{\mathrm{HMM}} = 
\begin{bmatrix}
    0.84 & 0.16 & 0    & 0   \\
    0    & 0.91 & 0.09 & 0   \\
    0    & 0    & 0.77 & 0.23\\
    0.04 & 0    & 0    & 0.96
\end{bmatrix} \;,
\end{equation}
where the rows and the columns correspond to the starting and destination states, respectively, in the following order: S1, systole, S2, diastole. Based on some set of observations (or emissions), the underlying hidden states can be estimated using the Viterbi algorithm \cite{Viterbi}.
However, if the Markov property is relaxed to be duration dependent, we get a so-called duration-dependent HMM (DHMM), also called as a hidden semi-Markov model (HSMM). This process requires the probability distribution of the durations for each state, usually modeled as a Gaussian distribution, and a modification of the state transition matrix to
\begin{equation}
\mathbf{A}_{\mathrm{HSMM}} =
\begin{bmatrix}
    0 & 1 & 0 & 0\\
    0 & 0 & 1 & 0\\
    0 & 0 & 0 & 1\\
    1 & 0 & 0 & 0
\end{bmatrix} \;,
\end{equation}
with the same row and column correspondence to the hidden states. Since the process is duration dependent, the onset and ending times of the underlying initial and final states could be outside of the measured observation sequence. This necessitates an extension to the Viterbi algorithm, which was described by Springer \textit{et al.} \cite{HSMM}. The introduction of time dependence was shown to be more accurate in terms of heart cycle segmentation. Both the original and duration-dependent models require a training phase with PCG signals, where the location of each heart sound is known. This can be a disadvantage if the training dataset is small, which would introduce biases to the segmentation model. However, training on a large dataset is not always possible because the heart sound labels are usually absent from fetal datasets, and accurately producing these labels requires manual labor. Although HSMM-based segmentation was originally not intended for use with fetal PCG analysis, it can be adapted to produce acceptable results by changing internal timing parameters such as the expected heart rate.
\begin{itemize}
    \item The duration-dependent HMM for heart cycle segmentation was first proposed by Schmidt \textit{et al.} \cite{Schmidt_2010}. Their paper compares the governing mathematics for both regular and duration-dependent models. In order to train the HSMM, a feature set is calculated from preprocessed PCG signals. The preprocessing is performed using a band-pass filter and a spike removal algorithm, which is done by removing certain time windows where the maximum absolute amplitude (MAA) is larger than three times the median of the MAAs. In the article different sets of features were compared based on their segmentation accuracy, the highest being a model only considering the homomorphic envelope. The observation probabilities for each state were modeled as a Gaussian distribution, calculated from the mean and covariance of the features. Similarly, the duration probabilities were modeled using a Gaussian function, with hard-coded mean and standard deviation values. With these distributions and the extracted features from an input signal, the hidden heart cycle states can be derived using the Viterbi algorithm.
    \item An extension to the HSMM method was proposed by Springer \textit{et al.} \cite{HSMM}. This extension came in two main forms, first by introducing the extended Viterbi algorithm for more accurate state decoding, and by implementing a more sophisticated observation probability estimation with logistic regression. This segmentation method is also referred to as LR-HSMM. The logistic regression used to create the observation probabilities is first trained on a set of extracted features, called an \textit{envelogram}. The envelogram, as mentioned in previous sections, contains the envelope of the signal calculated with Hilbert tranformation and its homomorphic filtered version, the local energy calculated using a sub-band of the power spectral density, and the absolute value of a selected DWT detail component. However, to get the conditional probabilities with the correct form, the logistic regression outputs have to be corrected for according to Bayes' theorem. Later, this LR-HSMM model was reimplemented and slightly improved in our previous work \cite{muller2024pypcg}, by extracting the duration distributions and other hard-coded values to parameters, in order to generalize the method for use with fetal cases.
    \item Lately, one of the most accurate methods using HSMM segmentation was introduced by Renna \textit{et al.} \cite{Renna-UNET}. They proposed a convolutional neural network (CNN) for heart cycle state estimation. The CNN architecture chosen was similar to the U-net design, originally used for image segmentation. The network described an encoder-decoder pair with skip connections between the corresponding convolution blocks to transfer local information. Each convolution block was connected via rectified linear unit activations, with max pooling layers between the encoding levels, upsampling layers between the decoding levels, and a final soft-max activation to get the final probabilities. The CNN was trained using the conventional preprocessing and envelogram features for HSMM, and originally used the PhysioNet 2016 Challenge data \cite{Physionet2016}. According to the article, the final output of the CNN could be transformed to heart cycle states in multiple ways. The simplest being taking the state with the maximum likelihood for each timestep with the constraint that only physiological state transitions can occur. The other methods used different HMMs (including HSMM) for a more robust segmentation. This method was reimplemented by Enériz \textit{et al.} \cite{Unet-impl} which was trained on the CirCor DigiScope dataset \cite{PhysioNet2022}. This was combined with our previous HSMM implementation and optimized for fetal data.
\end{itemize}
As it can be seen, most of Markov-model-based methods shown originally were not intended for processing fPCG. We still decided to include them due to their prominence in the literature discussing heart sound segmentation. For fine-tuning and parameter optimization details, see Chapter \ref{sec:results}.

With the recent increase in popularity of transformer models in neural network architectures, especially in language models, their prominence also increased for signal processing and PCG segmentation. Transformers make use of a mechanism called attention, where different parts of the sequence processed can influence and encode any other part of the sequence. The processing starts with tokenization, where the sequence is separated into smaller segments. In signal processing this can be achieved by time windowing. These segments, called tokens, are projected into a high-dimensional embedding space, where their coordinates encode different semantic meaning. The attention step takes these token embeddings as an input and transforms them so that the ``meaning" of a given token takes into the context it is located in. Mathematically a single \textit{head of attention} is
\begin{equation}
    \mathrm{Attention}(Q,K,V) = \mathrm{softmax} \left( \frac{QK^T}{\sqrt{d_k}} \right)V \;,
\end{equation}
where $Q$, $K$, and $V$ are query, key, and value matrices calculated from the embeddings and their respective weight matrices (usually notated as $W^Q$, $W^K$, and $W^V$), and $d_k$ is the dimensionality of the key vectors. For \textit{multiheaded attention} multiple of these calculations are performed in parallel each with their own set of weight matrices. These results are then concatenated and multiplied with a final output weight matrix ($W^O$), which gives us the output of a single ``attention block". To perform predictions with transformers, after each attention block a fully connected multilayer perceptron (MLP) is included. This pattern of attention and MLP blocks can be then repeated to refine the influence of each token to their neighbors.
\begin{itemize}
    \item Almadani \textit{et al.} used the transformer architecture in conjunction with U-net design to separate different sources in a mixed abdominal signal. The authors named this model \textit{FHSU-NETR} \cite{unet-transformer}. The model used three different embeddings and transformer models to separate maternal PCG and breathing signals from the fetal PCG. These transformers served as the encoder part of their respective U-nets and the skip connections were inserted after three repetitions of attention-MLP blocks. The U-net decoders used two convolutional layers with ReLU activations. The maternal PCG and respiration skip connections were subtracted from the respective fetal PCG layers using a \textit{tanh} activation.
    \item Kong \textit{et al.} proposed a hybridized model to segment the heart cycle, using transformer models and XGBoosted trees \cite{hybridized-classifier}. They made use of three different datasets: the 2016 PhysioNet challenge dataset, the data used by Springer \textit{et al.} to train and evaluate their model, and a simulated fetal dataset using a method proposed by Zemlyakov \textit{et al.} \cite{Zemlyakov_2020}. The segmentation process as described starts with time windowing followed by spectral feature extraction using Fourier Synchrosqueezed Transform. From these features only those which correspond to a frequency between 25 Hz and 200 Hz are kept. Using the selected frequency range both a transformer neural network and a random forest classifier with XGBoost is trained and their results are combined using a hybridized decision rule.
\end{itemize}

\begin{landscape}
\begin{table}[p]
\begin{center}
\caption{Summary of the discussed fPCG based fetal heart rate methods}
\begin{tabular}{l|l|l|l|l|l|l}
\hline\hline
\thead{FHR estimation method} & \thead{Dataset} & \thead{Preprocessing} & \thead{Detection} & \thead{Validation} & \thead{Accuracy} & \thead{Implementation} \\ \hline
Kovács \textit{et al.} (2000) \cite{FHR-rulebased} & Internal & \makecell{HP filter (20 Hz), LP filter (70 Hz)} & \gape{\makecell{20-40 Hz, 50-70 Hz peak detection,\\S1-S2 pair searching in time window}} & - & - & - \\
Várady \textit{et al.} (2003) \cite{FHR-advanced} & Internal & \gape{\makecell{Wavelet based inter-channel denoising \\(external recording available),\\BP filter (35-200 Hz)}} & \gape{\makecell{Envelope cross correlation,\\timing parameter estimation,\\state machine}} & - & Visual assessment & - \\
Chen \textit{et al.} (2006) \cite{FHR-rms} & Internal & \gape{\makecell{Analog LP filter (110 Hz),\\Digital HP filter (35 Hz),\\Simplified Spectral Subtraction}} & \gape{\makecell{RMS, threshold local maximum,\\peak merge, amplitude regularity}} & CTG & Visual assessment & \checkmark\textsuperscript{1} \\
Kósa \textit{et al.} (2011) \cite{FHR-heuristic} & Internal & - & \gape{\makecell{Multiple difference levels \\with smaller window sizes}} & Autocorrelation & Average Absolute Error & \checkmark\textsuperscript{1} \\
Zahorian \textit{et al.} (2012) \cite{FHR-zahorian} & Internal & \gape{\makecell{FIR filtering (several bands),\\Matched filtering}} & \gape{\makecell{Teager energy operator\\Autocorrelation\\``Merit" calculation}} & - & - & \checkmark\textsuperscript{1} \\
Yang \textit{et al.} (2014) \cite{FHR-mobile} & Internal & \gape{\makecell{Computational auditory scene analysis\\(external recording available),\\energy based rejection}} & Pattern matching & Doppler monitor & FHR difference & - \\
Tang \textit{et al.} (2016) \cite{FHR-cyclic} & \textit{simfpcgdb} & - & \gape{\makecell{Repetition frequency,\\cyclic frequency spectrum\\with windowing}} & \makecell{Multiple SNR levels,\\rule based and advanced method} & Tolerance based ratio & \checkmark \\
Dia \textit{et al.} (2019) \cite{Dia-FHR} & Internal & - & \Gape[6pt]{\makecell{Short-time Fourier transform,\\Non-negative matrix factorization,\\Moving median post-processing}} & CTG & \makecell{Correlation,\\Outlier ratio} & - \\
Huimin \& Xingyu (2020) \cite{FHR-LWT} & \textit{sufhsdb} & \gape{\makecell{EMD,\\Lifting wavelet denoising}} & \gape{\makecell{Hilbert transform,\\real cepstrum}} & - & - & - \\
Souriau \textit{et al.} (2023) \cite{FHR-bimodal-viterbi} & Internal & \Gape[6pt]{\makecell{ECG: HP filter (10 Hz),\\Band-stop filter (49-51 Hz),\\LP filter (80 Hz),\\Maternal ECG attenuation,\\PCG: BP filter (20-200 Hz),\\absolute value LP filter (15 Hz)}} & \gape{\makecell{Multimodal hidden Markov model\\with modified Viterbi algorithm}} & CTG & \gape{\makecell{Non-outlier ratio,\\Missing value ratio}} & - \\
Bhaskaran \& Arora (2024) \cite{bhaskaran-filterselect} & \textit{iiscfhsdb} & Comb filter (50 Hz) & \gape{\makecell{Multiple frequency bands,\\Hilbert envelope autocorrelation,\\Cyclic repetition frequency,\\decision rules}} & Database ground truth & \gape{\makecell{Mean absolute error,\\Positive predictive agreement,\\Ratio of valid FHR}} & - \\ \hline\hline
\end{tabular}
\label{tab:fhr-summary}
\end{center}
\textsuperscript{1}: Reimplemented by us, based on original paper for this evaluation
\end{table}
\end{landscape}

\section{Fetal heart rate estimation}
\begin{figure*}
    \centering
    \includegraphics[width=0.97\textwidth]{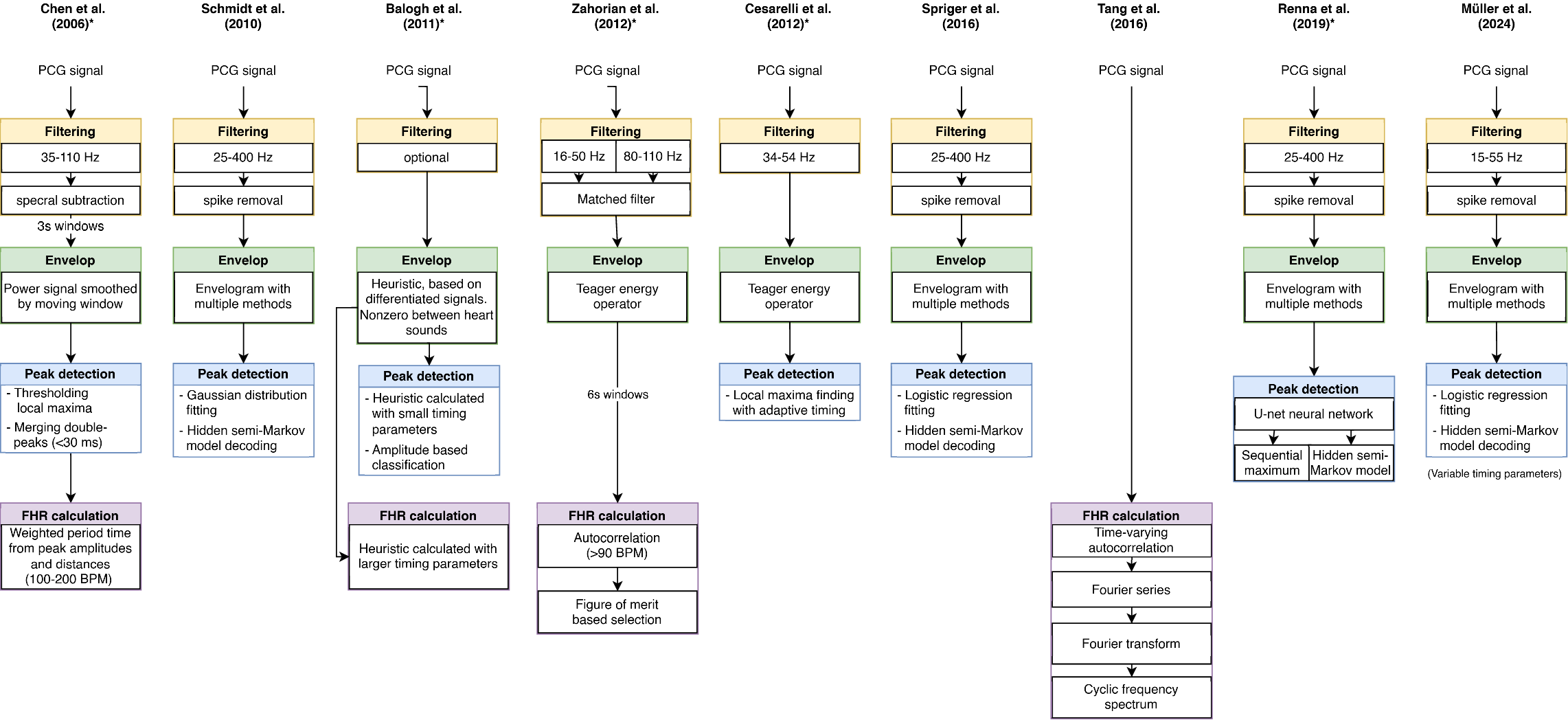}
    \caption{Simplified flowcharts of the evaluated fetal heart rate (FHR) estimation and fetal heart sound (fHS) detection methods. Reimplementations marked with an asterisk.}
    \label{fig:fhr-flow}
\end{figure*}

In this section fPCG based FHR estimation methods are described, summaries of these methods can be seen in Figure \ref{fig:fhr-flow} and Table \ref{tab:fhr-summary}.

A widely cited work is the rule-based method proposed by Kovács \textit{et al.} in 2000 \cite{FHR-rulebased}. This work served as a comparison base for most later FHR estimation methods. In this process, the raw fPCG signal is first filtered with a fifth-order high-pass filter at 20 Hz and a fourth-order low-pass filter at 70 Hz. After this preprocessing, the signal is divided into two frequency bands: 20-40 Hz and 50-70 Hz. The energies of these bands are calculated using a moving average, resulting in two signals where local maxima most often correspond to heart sounds. These energy signals are then encoded into rectangular pulses based on a threshold value calculated based on the previous encoded peaks. To calculate FHR based on these peaks, two types of timing patterns were fit on the final impulses: a two-peak and an eight-peak timing pattern. The timing patterns are constructed based on previous FHR predictions while also allowing slight deviations from the average and their correlation is calculated for the peaks. Based on the number of detected heart sounds the process can move on to confidence factor calculation or it can refine the timing patterns until the number of detections is acceptable. A confidence factor is also calculated from the encoded heart sound peak weights and previous confidence factor values. If the confidence was low, only the FHR was calculated without updating internal parameters, or at even lower confidence the FHR was not estimated. The algorithm was created to be implemented in an online processing paradigm, and the authors demonstrated an implementation on an 8-bit microcontroller, validating it with 80 ten-minute-long randomly selected CTG recordings. The authors analyzed the results and concluded that in approximately 90\% of the cases the curve remained between a 3 bpm tolerance level. In 5\% of the recordings the deviance was higher than the selected tolerance, but the deviation was under $\pm$5 bpm with a relatively lower confidence factor.

Várady \textit{et al.} implemented a real-time processing method for FHR in 2003 \cite{FHR-advanced} which included a secondary sensor to record external noise for later cancellation. They created a dataset consisting of 16 records from women between 28th and 40th week of gestation, with 9 of them having synchronous CTG recordings for verification. The noise canceling step was realized with a wavelet technique, decomposing both internal and external channels with coiflet-2 wavelets \cite{Strang_Nguyen_1997} to their sixth level. After an adaptive thresholding method a reverse wavelet transform reconstructed the noise free signal. An additional step to remove maternal influence in the signal was employed, realized with a band pass filtering between 35 Hz and 200 Hz. In the next step the envelope of the filtered signal was calculated with a unique method using the local minima and maxima of the signal. According to the authors this envelope represents well the amplitude dynamic of the signal while keeping the timings the same. To locate the amplitude impulses in the signal a cross-correlation score was calculated with a previously selected reference impulse along with an extended version of the algorithm by Kovács \textit{et al} \cite{FHR-rulebased}. With this modification the magnitude, the time location, and the probability of a burst were calculated which were used with a searching algorithm to locate S1 and S2 sounds. While the S2 could not be always located, given the time locations of the S1 sounds FHR could be calculated based on the time between each detection. With a concurrently calculated validity factor the accuracy was improved, by rejecting values with a low validity. The final implementation achieved an overall accuracy of 83\%, based on visual assessment on 9 CTG recordings.

In their work Chen \textit{et al.} proposed a method in 2006 to detect fHS and in turn estimate the fetal heart rate \cite{FHR-rms}. They made use of high- and low-pass filters to preprocess the signal and remove some noise outside the frequency range of the heart sounds. Another noise reduction step was applied, called spectral subtraction, which was originally used to enhance speech signals. Fetal heart rate was determined after detecting the heart sounds, which was achieved with an envelope peak detection algorithm. The envelope of the signal was calculated as the root mean square over a short time window. Then the peak detection was done with a simple local maximum search augmented with a global threshold. These detections were further filtered based on their regularity in amplitude and in time. The heart beat period was calculated from the difference of the detected peaks weighted by their amplitude. If a detection produced a predicted heart rate outside the expected range (100-200 bpm), it was discarded. As the authors described a high signal to noise ratio is required as well as the amplitude peaks to be high enough for an accurate prediction with the described method. To contextualize the accuracy of the predicted FHR, a confidence factor was introduced, which measured the regularity of the signal from the power spectral density of the envelope. Experiments with this method were conducted on 41 pregnant women with gestational age between 37 and 38 weeks. To validate the results a synchronized comparison was done with a CTG device. The authors concluded that the results are closely matching the CTG measurement and by labeling each time frame based on the confidence factor an easier evaluation can be done.

Kósa \textit{et al.} presented a heuristic method in 2008 for fetal heart sound detection and FHR calculation \cite{FHR-heuristic} which was then refined by Balogh and Kovács in 2011 \cite{Balogh-murmur}. According to the authors, this method is more robust to most types of noise present, and only relies on general features. The first step of both processes is calculating local intensity realized with windowed sums and their differences. Using the sum the local intensity and the signal a contrast enhancement was performed with a similar method. This produces a wavelike pattern and by taking only the positive values, individual heart cycles, or by choosing different window length parameters the individual heart sounds can be detected. Finally, if the results are acceptable to the conditions set by the user, further morphological analysis can be done on the signal, such as detecting splits, murmurs and extrasystole. The FHR estimation and beat detection was compared with an autocorrelation method, although no quantitative results were given, the described method achieved an average absolute error significantly lower than the reference. In terms of beat detection, the algorithm detected more beats and with a higher accuracy than the autocorrelation.

Zahorian \textit{et al.} \cite{FHR-zahorian} created an FHR estimation process with the aim to show that different fetal positions require different PCG frequency ranges for accurate measurements. They created a fetal heart monitoring system based on \cite{piezopolymer-sensor} by introducing a different amplifier and analog bandpass filters so that the user could change between two frequency ranges: 20-400 Hz and 80-400 Hz. After analog-digital conversion an additional user specified digital band pass filter was applied, with cutoff frequencies: 16-50 Hz or 80-110 Hz. Following this matched filtering was done on the signal with the aim to reduce the noise in the signal, the template for filtering calculated from the expected magnitude spectrum of an acceptable signal using an inverse Fourier transform. In the next step the Teager energy is calculated for the filtered signal. This energy operator produces an output similar to an envelope, and this is used in the next autocorrelation step. The autocorrelation used 6 second long time frames of the calculated energy, and from the local maxima of the result the periodicity of the signal could be estimated. In order to reject spurious peaks a lower bound for the heart rate was set at 90 bpm, and additional ``figure of merit" calculation was suggested. This figure of merit is calculated from previous merit scores and the apparent change in heart rate and at lower values the FHR estimate was rejected. The authors concluded that the proposed frequency bands were consistent with the clinical trial, where the fetal position corresponded well with the frequency setting of the analog band pass filters.

Yang \textit{et al.} demonstrated a method for FHR calculation with a mobile device for which they developed the hardware and the firmware \cite{FHR-mobile}. Their device had two sensors, one for recording the fPCG and another for recording the external noises. An initial noise cancellation step rejected time frames based on energy to reduce the motion artifacts in the recordings. With the additional external channel the authors could apply a special noise cancellation step to remove noises in the fPCG signal called \textit{computational auditory scene analysis} (CASA). In CASA denoising a gammatone filterbank is used to model the frequency selectivity of the human cochlea, resulting in a time-frequency representation. The authors used the interaural intensity difference of the ambient noise and the abdominal PCG recording to mask out certain values. Fetal heart rate was determined with a regression step, called \textit{adaptive matching} described in more detail in a previous article \cite{FHR-adaptivematching}. Adaptive matching was designed to overcome some of the shortcomings of the rule based method by Kovács \textit{et al.} They propose a different relation between the systolic time and the FHR by introducing a \textit{k} parameter as the ratio between the systolic time and a whole heart cycle. After an initial FHR estimation the parameter space for \textit{k} and FHR is scanned, to find the best fitting of the heart sound labels. The results of the mobile device were validated with a Doppler monitor, on 8 pregnancies between the gestational age of 37 and 40 weeks. The described method and device had similar FHR values to the results given by the Doppler monitor, with only around a 10\% error.

Tang \textit{et al.} published a FHR calculation method using the repetition frequency of the heart sounds \cite{FHR-cyclic}. According to the article, their method did not employ any preprocessing steps. The main component of the described method is a cyclic spectral density calculation. First a time-varying autocorrelation is calculated, which captures the local regularity of the signal. The Fourier series expansion based on the signal time transforms the autocorrelation to the cyclic correlation function and introduces the cyclic frequency as parameter. Taking the Fourier transform for the other dimension, results in the cyclic spectral density of the signal which is used to calculate the cyclic frequency spectrum. The periodicity is accurately captured in this spectrum, and the FHR can be obtained from its dominant peak. The method described this way only captures an average heartrate and by modifying the method with a short time Fourier transform an FHR curve can be estimated for further evaluation. This method was tested with multiple levels of signal to noise ratios (SNR), and for validation the rule-based method by Kovács \textit{et al.} \cite{FHR-rulebased} and the advanced method by Várady \textit{et al.} \cite{FHR-advanced} was used. For the best SNRs the accuracy of the proposed method was generally lower, but the other methods lost their accuracy significantly at lower SNR values. In the worst case scenario, with an SNR of -26.7 dB, an accuracy of 81\% was achieved, while the other two methods missed too many heart sounds and their accuracy was not evaluated.

Non-negative matrix factorization based FHR estimation was proposed by Dia \textit{et al.} \cite{Dia-FHR}. In their article the authors try to leverage the quasi-periodic nature of PCG signals. A short-time Fourier transform using a long window size (4 seconds) is used to capture multiple heartcycles and estimate their periodicity. This spectrogram can be modeled as a time-varying \textit{source-filter model}, in this case a Dirac comb with fundamental frequencies at each possible heart rate (30-300 bpm). The model separates the signal into two terms, the \textit{excitation} and \textit{filter}, which were further decomposed with non-negative matrix factorization (NMF or NNMF). Performing NMF on the excitation part of the STFT spectrum gives a spectral and a temporal term. By selecting the original filter corresponding to the maximum temporal part the best fitting heart rate can be obtained based on its fundamental frequency. However, this process can also select the second harmonic or the half sub-harmonic. To fix this problem the authors implemented a post-processing method based on the derivative of the estimate and a moving median filter. The method was tested using PCG and CTG recordings from four volunteers, although, not all of the recordings were usable. Correlation of the estimate and two of the usable CTG signals were calculated which gave 91\% and 84\%. Outlier percentage was also calculated for all volunteers, which was on average 8.4\%, where an outlier was considered if the estimate differed from the reference median FHR by more than 10\%. 

Huimin and Xingyu decribed a unique way to determine the FHR based on a cepstrum method \cite{FHR-LWT}. The process included a denoising step with EMD and wavelet transform, acquired with the lifting wavelet technique. From the EMD step the first four IMFs with the highest frequencies were selected for the wavelet based denoising using a ``semi-soft" threshold \cite{thresholding}, the rest were not processed this way. The denoised signal was used to estimate the FHR, achieved by calculating the real cepstrum of its Hilbert envelope. According to the authors since the fPCG is a periodic signal, a series of pulses should appear in the cepstrum. Using these pulses the heart rate can be obtained by detecting the location of the maximum cepstrum value between 0.2 seconds and 1 second. The described method was tested on 20 randomly chosen recordings from the Shiraz University dataset, although the dataset is only refered to as ``PhysioNet database". No quantitative results were shown but the authors concluded that the method could accurately estimate the FHR.

Souriau \textit{et al.} developed a hybrid method for FHR estimation using both fPCG and fetal electrocardiography (fECG) \cite{FHR-bimodal-viterbi}. The different modalities were processed with different filters to remove noise from the signals. In the fECG recording the lower frequencies and the powerline noise was removed using a high-pass filter at 10 Hz and a band-stop filter between 49 and 51 Hz, respectively. While the fPCG was preprocessed using a band-pass filter with 20 and 200 Hz cutoff frequencies. The fPCG was further processed by taking its absolute value and passing it through a low-pass filter with 15 Hz cutoff frequency. Two estimates for the FHR were calculated initially with a monomodal setup, using the NMF method described by Dia \textit{et al.} \cite{Dia-FHR}. Their main contribution was by introducing a multimodal hidden Markov model to refine the FHR estimates by combining information from both types of signals. A modified Viterbi algorithm was also suggested to further improve the accuracy and reduce the ``maternal confusion", which means to reduce sections where the maternal heart rate is given as the FHR. Evaluation was done using data from 6 pregnancies with CTG records as comparison and maternal ECG recordings to determine maternal confusion of the method. The authors introduced the non-outlier ratio, where points with lower than 12.5 bpm difference were counted and divided by the total amount of FHR points. The modified Viterbi algorithm lowered maternal confusion while in multiple cases improved FHR non-outlier ratio.

Bhaskaran and Arora published a method which used multiple frequency bands and combined autocorrelation and cyclic frequency calculations \cite{bhaskaran-filterselect}. To reduce the amount of powerline noise in the recordings a comb filter was used with a 50 Hz base frequency. Then the signal was analyzed at different frequency bands between 10 and 200 Hz, with different bandwidths which were between 30 and 190 Hz. Two features were calculated using the Hilbert envelope autocorrelation (HAC) and the cyclic repetition frequency (CRF). For both cases, the local maxima were used to derive the actual features by including other properties such as the PSD and the number of peaks. A given frequency band was selected based on the values of the features to calculate an initial FHR estimate. A complex decision rule was then used to determine to further update the estimate or reinitialize the process by another frequency band selection. This estimation, refinement, reinitialization process was performed for each 4 seconds of the input. The method was evaluated on the Indian institute dataset by calculating the mean absolute error, the positive predictive agreement, and the ratio of valid FHR outputs. The HAC feature gave a 6.97 bpm error, 88\% agreement, and 92.7\% valid FHR, while the CRF feature resulted in 6.87 bpm error, 89\% agreement, and 90.4\% valid FHR for all signals.

\section{Methods and materials}
The methods we tested can be categorized into two main groups: FHR estimation and fHS detection. Based on our experience with the literature one important task is to create a common testing environment, so that the results are comparable to each other. For this we used a dataset where the signals were selected from the recordings created by Ferenc Kovács. Based on a preliminary signal quality selection criteria, 50 one-minute-long sections were selected. This dataset can be then further augmented with additional sections in the future. These selected sections were manually labeled for the first and second heart sounds based on the apparent energy, frequency, and time delay of the observed impulses. As far as we know this is currently the largest fPCG dataset with labeled individual heart sounds, containing 6758 S1 and 6729 S2 manual labels. The previously mentioned datasets all have their drawbacks for determining fHS detection accuracy, or even FHR estimation accuracy. To our knowledge only the Shiraz and Indian Institute datasets contain FHR data, and only the Shiraz dataset contains multiple FHR values for a given signal. However due to its recording setup, the signals in that dataset do not follow the commonly observed frequency ranges and distributions, making frequency dependent fPCG methods (which there are several) unsuitable. There is another possibility, which is to manually label signals or selected sections from these standardized datasets. Since for previous work we already had to create a dataset, this was not done and we decided to use our own dataset \cite{muller2024pypcg}. Our dataset contained 50 records, which were 60 seconds long, as mentioned previously. The signals were recorded with the Fetaphon device, which operates with 333 Hz sampling frequency and 8 bits of precision.

The comparison of detection methods were done in the same way as in our previous works \cite{muller2024pypcg}, which is based on the evaluation by Renna \textit{et al.} In this benchmarking method we transform the problem to a classification accuracy measurement by creating so called ``tolerance intervals" around a given manual label. If the detection lies inside this interval it is considered a true positive (TP), otherwise it is counted as a false negative (FN). False positives (FP) can also be calculated if the tolerance interval is measured around the detection and a ground truth label is not found in that region. This transformation allows the use of common accuracy measures such as the positive predictive value (PPV) and F1-score. These measures are calculated using the following formulae:
\begin{equation}
    \mathrm{PPV} = \frac{TP}{TP+FP},
\end{equation}
\begin{equation}
    \mathrm{F1} = \frac{2 \times PPV \times TPR}{PPV + TPR},
\end{equation}
where $TPR$ is calculated using all positive cases (P) and true positives:
\begin{equation}
    \mathrm{TPR} = \frac{TP}{P}
\end{equation}
However, these measures heavily depend on the chosen size of the tolerance interval. By varying the tolerance we can measure the dependence of the accuracies on the tolerance, this relation can be visualized as a graph that we decided to call \textit{Score-vs-Tolerance}. These Score-vs-Tolerance plots visually resemble the receiver operating characteristic curves, but the background process and thus the meaning is different. In Score-vs-Tolerance plots there are two important properties we can observe: the \textit{rise} which shows the robustness or precision of the detections, and the \textit{plateau} which shows the theoretical maximum score.

Another comparison measure we previously introduced is based on the word error rate \cite{word-error-rate}, originally used in voice recognition assessment. In it three types of errors are calculated: insertions (INS), deletions (DEL), and substitutions (SUB). Similarly to the previous method, a tolerance interval was considered around each manual label to quantize the detection error. Insertion errors were counted if there were additional detections between the tolerance intervals, deletions were counted if there were no detections inside the interval, and substitution was counted if the detection was the wrong type (for example: S1 instead of S2). These errors were then separately divided by the total amount of manual labels to get a percentage value.

\begin{figure}
    \centering
    \includegraphics[width=0.9\columnwidth]{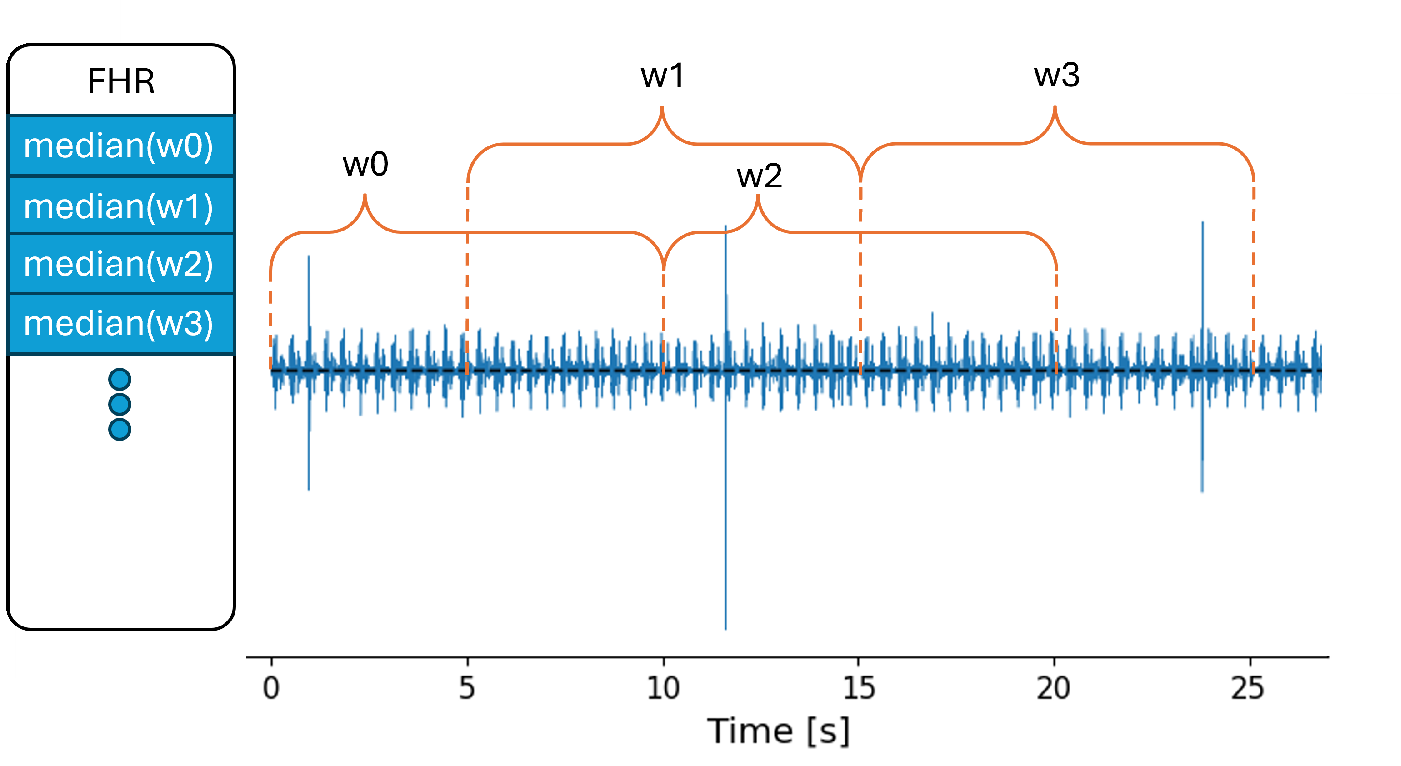}
    \caption{Visualization of fetal heart rate (FHR) calculation from heart sound detections. Ten-second-long time windows labeled as w0-w3 with 50\% overlap, on a sample fPCG signal. The median S1-S1 times are collected into the FHR array}
    \label{fig:fhr-calc}
\end{figure}

Benchmarking FHR estimation methods using our data required the calculation of ground truth FHR values based on our labels. To achieve this, the signals were cut into 10 second long windows with 50\% of overlap, and the heart rate was measured based on the median S1-S1 time in a given window, illustrated in \ref{fig:fhr-calc}. To reject outlier values due to not labeling noisy regions, a plausible S1-S1 time range was introduced. Meaning if an individual time difference corresponded to a heart rate value above 210 bpm or below 80 bpm, it was rejected. This range was taken from Tang \textit{et al.} \cite{FHR-cyclic} and all FHR estimation methods, where it was applicable, parameters were changed to expect the heart rate in this range. Other FHR estimation methods in the literature specified similar ranges, and this was chosen because it contained both the physiological and pathological FHR values. The parameters describing the expected FHR were extracted to be easily configurable. The previously mentioned way of calculating FHR based on heart sound labels was also used to transform the outputs of the fHS detection methods. FHR estimation accuracy calculation was done using the mean square error (MSE), which is a commonly used way to calculate similarity between pointwise similarity between signals. After calculating MSE for all records several statistical measures were taken: the mean, standard deviation (SD), minimum, maximum, and interquartile range (IQR)

Benchmark results were calculated for almost all methods where implementation (or reimplementation) is available. These are marked with a check mark in Table \ref{tab:fhr-summary} and Table \ref{tab:fhs-summary}.

\section{Results}
\label{sec:results}
We will refer to the methods by the first author of the paper where they were described. In case of Renna \textit{et al.} the temporal modeling was done in two ways, sequential maximum (Seqmax) and HSMM (using the pyPCG implementation), these are referred to as Renna-Seqmax and Renna-HSMM, respectively.
In case of HSMM methods (Schmidt, Springer, Müller, Renna-HSMM) a 10-fold cross validation was performed using the same dataset. The CNN models also require training, however, a larger dataset is needed to avoid overfitting. We used the CirCor DigiScope dataset for initial training, and transfer learning was performed using our fetal data. This was achieved by lowering the learning rate while re-training.

\begin{table*}[t]
\setlength{\tabcolsep}{18pt}
\caption{Accuracy measures of the tested methods on S1 and S2 detection separately}
\begin{center}
\begin{tabular}{ll|ll|ll|ll}
\hline\hline
\multicolumn{2}{l|}{\multirow{2}{*}{Method}} & \multicolumn{2}{c|}{\gape{PPV (\%)}} & \multicolumn{2}{c|}{F1 (\%)} & \multicolumn{2}{c}{MAE (ms)} \\
\multicolumn{2}{l|}{}                        & \multicolumn{1}{l|}{S1}  & S2 & \multicolumn{1}{l|}{S1} & S2 & \multicolumn{1}{l|}{S1}  & S2 \\ \hline
\multicolumn{2}{l|}{\gape{Müller}}                  & \multicolumn{1}{l|}{\textbf{97.6}} & 86.7 & \multicolumn{1}{l|}{\textbf{97.4}} & 86.8 & \multicolumn{1}{l|}{\textbf{12.2 $\pm$ 8.0}}  & 19.8 $\pm$ 13.7 \\
\multicolumn{2}{l|}{Springer}                & \multicolumn{1}{l|}{76.9} & 74.5 & \multicolumn{1}{l|}{76.9} & 74.6 & \multicolumn{1}{l|}{23.8 $\pm$ 11.8} & 30.2 $\pm$ 19.1 \\
\multicolumn{2}{l|}{Cesarelli}               & \multicolumn{1}{l|}{79.8} & N/A  & \multicolumn{1}{l|}{79.7} & N/A  & \multicolumn{1}{l|}{27.4 $\pm$ 28.0} & N/A \\
\multicolumn{2}{l|}{Balogh}                  & \multicolumn{1}{l|}{94.6} & 79.6 & \multicolumn{1}{l|}{94.9} & 78.5 & \multicolumn{1}{l|}{15.4 $\pm$ 8.2}  & 39.5 $\pm$ 37.3 \\
\multicolumn{2}{l|}{Schmidt}                 & \multicolumn{1}{l|}{95.5} & 86.7 & \multicolumn{1}{l|}{95.6} & 86.9 & \multicolumn{1}{l|}{14.1 $\pm$ 12.1} & 25.2 $\pm$ 23.8 \\
\multicolumn{2}{l|}{Chen}                 & \multicolumn{1}{l|}{28.9} & N/A & \multicolumn{1}{l|}{29.0} & N/A & \multicolumn{1}{l|}{62.0 $\pm$ 13.5} & N/A \\
\multirow{2}{*}{Renna}    & Seqmax   & \multicolumn{1}{l|}{94.3} & 89.6 & \multicolumn{1}{l|}{87.3} & 83.1 & \multicolumn{1}{l|}{17.0 $\pm$ 26.2} & 17.9 $\pm$ 8.8 \\ 
                                  & HSMM     & \multicolumn{1}{l|}{84.5} & \textbf{91.4} & \multicolumn{1}{l|}{85.4} & \textbf{91.3} & \multicolumn{1}{l|}{33.9 $\pm$ 52.8} & \textbf{17.3 $\pm$ 12.2} \\ \hline\hline
\end{tabular}
\label{tab:results-12}
\end{center}
Positive predictive value (PPV), F1-score (F1), and mean average error (MAE) shown. For PPV and F1 a constant tolerance of 30 ms was used. Best values highlighted in bold.
\end{table*}

\begin{figure}
    \centering
    \includegraphics[width=0.9\columnwidth]{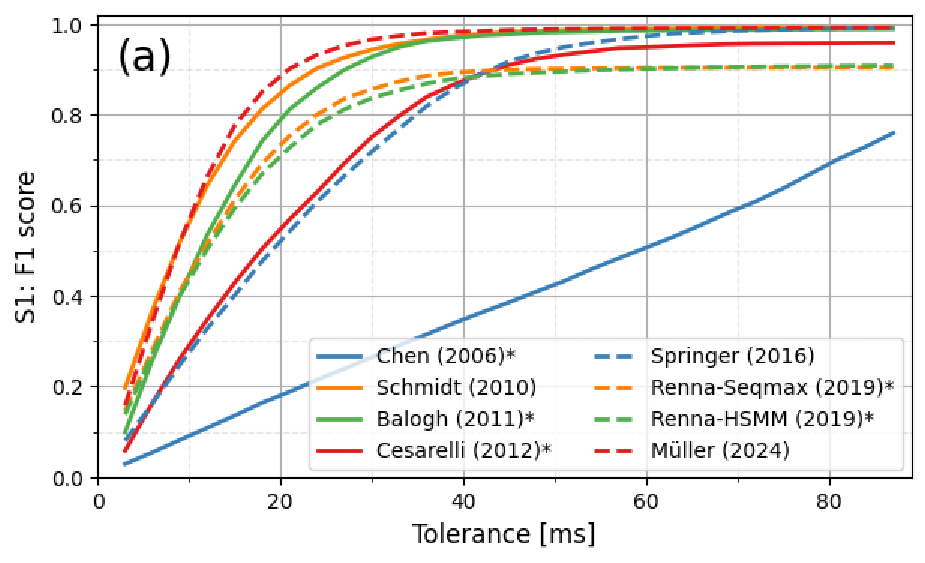}
    \includegraphics[width=0.9\columnwidth]{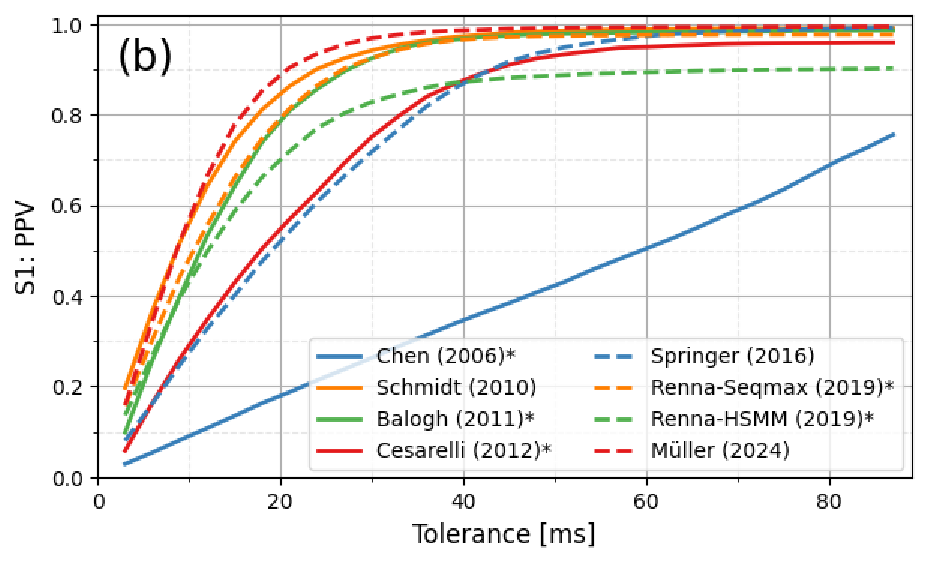}
    \caption{Fetal S1 Score-vs-Tolerance plots. (a) F1-score (b) Positive predictive value (PPV). Reimplementations marked with an asterisk}
    \label{fig:s1-svt}
\end{figure}

\begin{figure}
    \centering
    \includegraphics[width=0.9\columnwidth]{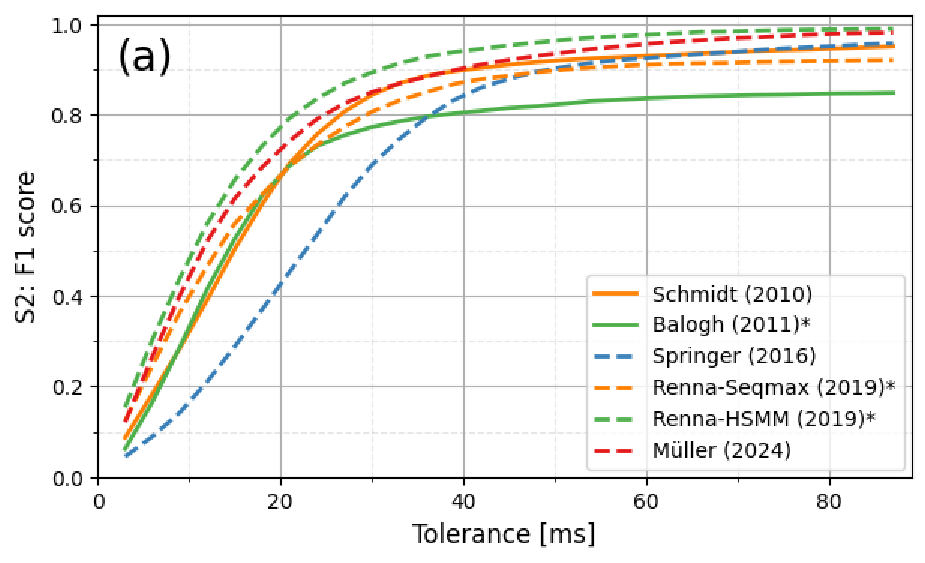}
    \includegraphics[width=0.9\columnwidth]{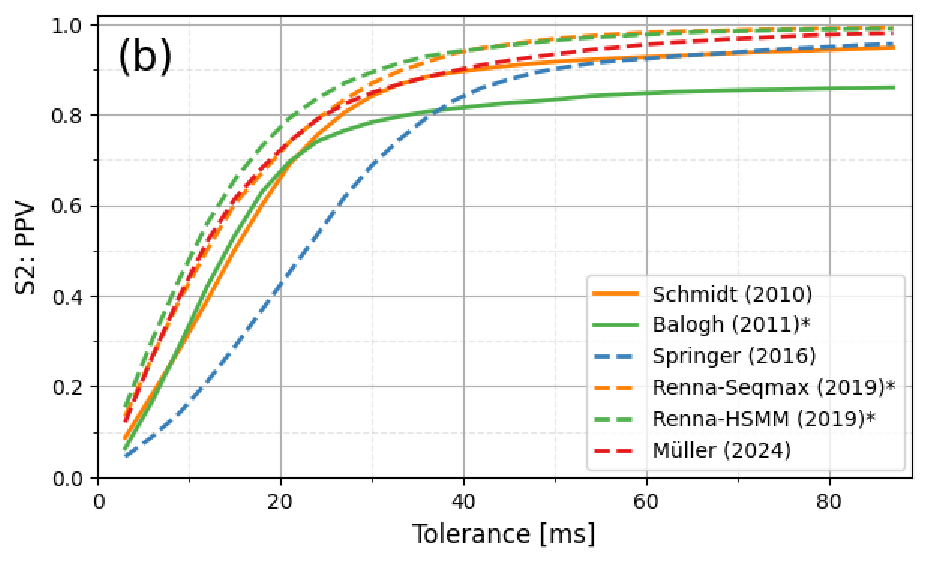}
    \caption{Fetal S2 Score-vs-Tolerance plots. (a) F1-score (b) Positive predictive value (PPV). Reimplementations marked with an asterisk}
    \label{fig:s2-svt}
\end{figure}

Results for heart sound detection accuracy measures can be seen in Table \ref{tab:results-12}, with the error rates in Table \ref{tab:results-wer}, both evaluated with a constant tolerance of 30 ms. Score-vs-Tolerance plots are provided for S1 and S2 separately in Figure \ref{fig:s1-svt} and Figure \ref{fig:s2-svt}. Our previous HSMM implementation with logistic regression (Müller) achieved the best accuracy scores for S1 detection (PPV: 97.1\%, F1: 97.4\%, MAE: 12.2 $\pm$ 8.0 ms) with a comparably good performing S2 detection. This pattern is also seen in the error rates for this method as it received the lowest sum error rate of 17.8\%, and in the Score-vs-Tolerance plot in Figure \ref{fig:s1-svt} where it had both a sharp rise and a high plateau. The best method in terms of S2 detection accuracy is the Renna-HSMM method (PPV: 91.4\%, F1: 91.3\%, MAE: 17.3 $\pm$ 12.2 ms), however, this is not reflected in the error rates, mainly due to the S1 inaccuracies weakening these scores. Although training it on a purposely built fPCG database it could be one of the most robust methods for detection, based on general experience with neural networks. In Figure \ref{fig:s2-svt} a gradual rise can be observed and a slightly sloped plateau for this almost all methods meaning that both the precision and the theoretical maximum score was lowered in the S2 detection task. It is important to mention the cause of the significantly decreased accuracy of the Chen method. As it can be seen, the S1 mean absolute error had low variance across the records ($\pm$13.5 ms) suggesting that most likely a constant delay was introduced and not corrected for. The error rates further support this delay hypothesis, as the insertion and deletion scores are similar implying that the detections consistently lied outside of the tolerance interval. This could be caused by a misinterpretation of the algorithm in the reimplementation. However, by observing the relevant Score-vs-Tolerance plots it can be seen that this is most likely not the case. Since a delay would be seen as a delayed rise in the score.

\begin{figure*}[!h]
    \centering
    \includegraphics[width=\textwidth]{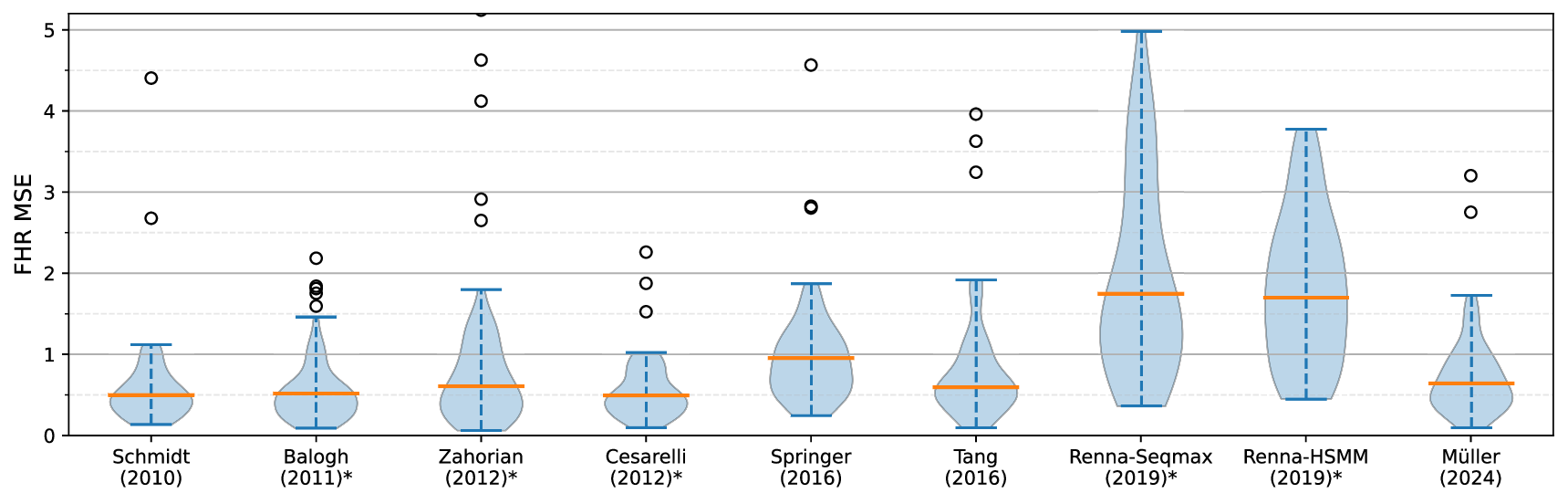}
    \caption{Fetal heart rate mean square error (MSE) violin plots of the tested methods, orange lines mark the median, empty circles mark the outliers, blue lines mark the first and the third quartile. The number of outliers not shown (MSE larger than 5) for each method: Schmidt: 2, Zahorian: 3, Cesarelli: 2, Springer: 2, Tang: 4, Renna-Seqmax: 5, Renna-HSMM: 3, Müller: 2. Chen not plotted because of large amount of outliers. Results marked with an asterisk are based on reimplementations}
    \label{fig:fhr-error-1}
\end{figure*}


Heart rate estimation accuracy is shown in Table \ref{tab:results-fhr} with a visualization shown in Figure \ref{fig:fhr-error-1}. Interestingly, the high accuracy of the heart sound detection algorithms did not translate to a highly accurate FHR estimation. As it can be seen, the best mean MSE was achieved by the Balogh method with the lowest worst case MSE and a relatively low IQR. Another important thing to note is the substantially outlying worst case MSE for the Cesarelli method. This was caused by a single case where the estimated FHR was close to the expected minimum ($\approx$ 90 bpm).

\begin{table}[b]
\setlength{\tabcolsep}{6.5pt}
\caption{Error rates of the tested methods}
\begin{center}
\begin{tabular}{ll|l|l|l|l}
\hline\hline
\multicolumn{2}{l|}{\gape{Method}} & \multicolumn{1}{c|}{INS (\%)} & \multicolumn{1}{c|}{DEL (\%)} & \multicolumn{1}{c|}{SUB (\%)} & \multicolumn{1}{c}{SUM (\%)} \\ \hline
\multicolumn{2}{l|}{\gape{Müller}}                  & \textbf{9.0} & 8.6 & \textbf{0.1} & \textbf{17.8} \\
\multicolumn{2}{l|}{Springer}                & 28.7 & 28.9 & \textbf{0.1} & 57.6 \\
\multicolumn{2}{l|}{Cesarelli}               & 12.2 & 10.5  & 1.6 & 24.3 \\
\multicolumn{2}{l|}{Balogh}                  & 14.2 & \textbf{7.8} & 6.1 & 28.0 \\
\multicolumn{2}{l|}{Schmidt}                 & 10.1 & 10.2 & 0.4 & 20.6 \\
\multicolumn{2}{l|}{Chen}                    & 35.8 & 36.1 & 0.2 & 72.0 \\
\multirow{2}{*}{Renna}  & Seqmax         & 21.3 & 8.4 & 0.7 & 30.4  \\
                        & HSMM           & 13.8 & 12.8 & 2.1 & 28.8  \\ \hline\hline
\end{tabular}
\label{tab:results-wer}
\end{center}
Insertion (INS), deletion (DEL), and substitution (SUB) errors shown as percentages separately as well as their sum. A constant tolerance of 30 ms was used. Best values highlighted in bold.
\end{table}

\begin{table}[b]
\setlength{\tabcolsep}{8pt}
\caption{Fetal heart rate mean square error (MSE) values of the tested methods}
\begin{center}
\begin{tabular}{ll|l|l|l|l|l}
\hline\hline
\multicolumn{2}{l|}{Method} & \multicolumn{1}{c|}{Mean} & \multicolumn{1}{c|}{SD} & \multicolumn{1}{c|}{Min} & \multicolumn{1}{c|}{Max} & \multicolumn{1}{c}{IQR} \\ \hline
\multicolumn{2}{l|}{Müller}                  & 1.380 & 3.441 & 0.098 & 24.15 & 0.583 \\
\multicolumn{2}{l|}{Springer}                & 1.879 & 3.734 & 0.245 & 22.73 & 0.702 \\
\multicolumn{2}{l|}{Cesarelli}               & 39.24 & 266.5 & 0.100 & 1886 & \textbf{0.463} \\
\multicolumn{2}{l|}{Balogh}                  & \textbf{0.644} & \textbf{0.500} & 0.093 & \textbf{2.185} & 0.499 \\
\multicolumn{2}{l|}{Schmidt}                 & 0.865 & 1.268 & 0.137 & 6.332 & 0.395 \\
\multicolumn{2}{l|}{Chen}                    & 31.03 & 50.67 & 1.004 & 261.3 & 24.85 \\
\multirow{2}{*}{Renna}  & Seqmax                  & 9.442 & 33.48 & 0.361 & 220.6 & 2.118 \\
                        & HSMM                  & 4.152 & 11.13 & 0.450 & 63.65 & 1.525 \\
\multicolumn{2}{l|}{Tang}                    & 1.550 & 2.379 & 0.099 & 12.52 & 0.704 \\
\multicolumn{2}{l|}{Zahorian}                & 1.482 & 2.765 & \textbf{0.062} & 17.71 & 0.851 \\ \hline\hline
\end{tabular}
\label{tab:results-fhr}
\end{center}
Mean, standard deviation (SD), minimum, maximum and interquartile ranges (IQR) calculated based on MSE values for all signals in the testing data.  Best values highlighted in bold.
\end{table}

\section{Conclusions}
We demonstrated a common platform to compare different fetal heart sound detection and fetal heart rate estimation methods. Based on the original papers multiple of these were reimplemented and made openly available, however, the accuracy of these implementations may not be representative of the original algorithms. Our results suggest that there are aspects where each method performs the best. For S1 detection our implementation of the LR-HSMM, for S2 detection the Renna CNN model with HSMM temporal modeling, and for FHR estimation the Balogh heuristic method were the best.

We see several important problems and open questions which need to be addressed in the future, such as: the lack of large labeled fPCG datasets both in terms of heart sound and fetal developmental disorder labels, most datasets do not contain metadata about factors which can cause complications during pregnancy or labor, the datasets which include important metadata (CHD, complications etc.) are not public, almost all fPCG processing research is about just determining the FHR which was shown to have low correlation with actual problems \cite{CTG-continuous, CTG-antenatal}, FHR estimation methods are usually not evaluated on the same datasets and in the same manners, which can cause promising methods to appear unfavorable, and no long-term monitoring data is available.

With this article we aim to raise awareness for these problems and suggest a standardization method for evaluation and comparison of heart sound detection and FHR estimation methods.

\section*{Acknowledgment}
The evaluation data was selected from a larger dataset provided by Prof. Ferenc Kovács.

We would like to thank David Springer, Daniel Enériz, Ivan Vican, Hong Tang, Lingping Kong and their teams for making their implementation source code available.

The work was supported by the  TKP2021-NVA-27 grant, funded by the Ministry of Innovation and Technology with support from the National Research Development and Innovation Fund under the TKP2021 program.

\bibliography{refs}
\bibliographystyle{ieeetr}

\end{document}